\documentstyle[11pt,aaspp,tighten]{article}
\begin{document}

\title{Modes of Elliptical Galaxies}

\author{Ue-Li Pen}

\affil{Princeton University Observatory \\
	Princeton, NJ08544 \\
	I: upen@astro.princeton.edu}

\date{March 27, 1993}

\newcommand{\myref}[1]{($\!\!$~\ref{#1})}

\begin{abstract}

Long lived modes of elliptical galaxies can exist {\it \`a la} van
Kampen.  Specific systems may possess long lived oscillations which
Landau damp on time scales longer than a Hubble time.  Some physical
processes such as a close encounter, tidal forces from a cluster or an
orbiting satellite could preferentially excite a coherent mode.  These
may relate to the observed faint structure in elliptical galaxies such
as shells and ripples.  Their detection in projected phase space would
ultimately provide a detailed probe of the underlying potential.

I give an overview of linear perturbations to stationary solutions of
the Vlasov equation, including a discretized Hermite polynomial
expansion which explicitly demonstrates completeness and orthogonality
of solutions.  Some exact solutions are shown, which implies the
feasibility of such a procedure and suggest future fully numerical
studies.

\end{abstract}
\section{Introduction}

Elliptical galaxies provide a clean laboratory to study the physics of
self-gravitating fluids.  We would like to study the possibility of
applying normal mode analysis to complement the observational data.
The conventional wisdom is that most excitations Landau damp or phase
mix at rates comparable to the dynamical time, making it impossible to
observe the equivalent of p-modes in stars.  Exceptions to this rule
have been studied by previous authors.  Mathur (1990) has shown that
not all modes decay by proving the existence of discrete eigenvalues
that allow stable oscillations.  Weinberg (1991) numerically studied
these modes in the vertical structure of the disk.  A different
approach was shown in (Weinberg 1993) where he numerically studied
modes which decay on time scales long compared to the dynamical time.
In this article I wish to extend this search by examining qualitative
relationships between various modes in linear theory.

I follow the van Kampen approach to construct these modes.  Since Case
(1959) has shown van Kampen's method to be equivalent to Landau's
treatment, we can use the qualitative features of van Kampen modes to
infer properties of Landau damping.  In the Jeans analysis, for
example, the normal van Kampen modes are singular (Binney and
Tremaine 1987, appendix 5A).  This not only makes them impossible to
excite, but also violates the Landau damping assumptions, where the
distribution must be differentiable for $v$.  Conversely, I will
suggest how non-singular modes might be constructed, and show explicit
examples in some one dimensional toy systems.  A differentiable van
Kampen mode implies the existence of a mode which does not Landau
damp.  If the system is furthermore proven to be stable, we know that
the Landau integration contour must contain a pole on the real axis, without
actually having to find it.

In this paper I first give a general treatment of linear perturbations
to review the notation and physical interpretation.  Next I will
assume spherical symmetry to study the modes found in simple
stationary systems.  While they are too simple to be applicable to
real astrophysical systems, they demonstrate how non-singular
van Kampen modes might exist.   Following that, I will
construct a frame work based on matrix expansions, which could be used
to study modes numerically.  This method is applied to the example of
the Jeans' instability.  We then construct an explicitly self-adjoint
complete representation of the linear Vlasov operator in the presence
of discontinuous functions.  By showing how an
unbounded operator which is not explicitly self-adjoint can possess a
complete set of orthogonal eigenfunctions we find a resolution to
Case's puzzle.  The argument also lends strength to the feasibility of
a general mode decomposition.  The paper concludes with some
speculation on possible astrophysical implications and suggestions for
future work.

\section{Formulation}

Qualitatively, one proceeds as follows: we consider a stable
stationary system (such as an elliptical galaxy), and look for
periodic perturbations.  This is accomplished by first constructing a
small perturbation consisting of a smooth periodic pattern in the
background potential, i.e. a periodic hamiltonian flow.  To linear
order, this pattern does not interact with itself.  It does, however,
effect the background fluid.  Having assumed the system stable, we
know that no resonant growth can result.  We need only study the
response of the background, and we are done.  If we are unfortunate,
the background response may cancel the perturbation exactly, and such
an example is given in the next section.  Or the background response
to a periodic perturbation may be chaotic, for example if the
potential is not integrable.  In general, however, we have some
non-trivial background response, as constructed explicitly in
section~\ref{jeanssec} for the case of the Jean's instability.

To phrase this  mathematically, we are interested in the solution to
Vlasov's equation
\newcommand{\pdx}[2]{\frac{\partial #1}{\partial #2}}
\begin{equation}
\pdx{f}{t} + \sum_{i=1}^3 \left[ v_i \pdx{f}{x_i}
	- \pdx{\Phi}{x_i}\pdx{f}{v_i} \right] = 0
\label{vlasov}
\end{equation}
\begin{equation}
\nabla^2 \Phi=4 \pi G \rho, \ \ \ \rho(\vec{x})=\int f d^3v,\ \ \
\vec{x}=(x_1,x_2,x_3), \ \ \ f=f(\vec{x},\vec{v};t) . \nonumber
\end{equation}
The problem under consideration consists of a stationary background
solution $f_b(\vec{x},\vec{v})$ and a periodic perturbation
$f_p(\vec{x},\vec{v})$ such that $f=f_b+\epsilon e^{i\omega t}f_p$,
where $\epsilon$ is a small number.
The sufficient condition for the perturbation equation to be valid is that
\begin{equation}
\epsilon\pdx{\Phi_p}{x_i}\ll\pdx{\Phi_b}{x_i}
\label{ineq}
\end{equation}
at all points $\vec{x}$ for $i=1,2,3$, which is weaker than requiring
$\epsilon f_p < f_b$ at all points in phase space.  The latter
condition only becomes necessary when we consider negative
perturbations, but positive perturbations are allowed even when
singular in velocity space.

To first order in $\epsilon$ the perturbation satisfies the linear
equation
\newcommand{\bL}{{\bf L}}
\newcommand{\bLa}{{\bf L}_1}
\newcommand{\bLb}{{\bf L}_2}
\begin{eqnarray}
\bL f_p &=& \omega f_p, \ \ \ \ \ \ \  \bL=\bLa+\bLb,\nonumber\\
\bLa&=&i \sum_{i=1}^3 \left[ v_i \pdx{}{x_i}
	- \pdx{\Phi_b}{x_i}\pdx{}{v_i} \right], \nonumber\\
\bLb&=& - 4i\pi G \sum_{i=1}^3  \pdx{f_b}{v_i} \pdx{}{x_i}
\nabla^{-2} \int d^3v .
\label{lvlasov}
\end{eqnarray}

This decomposition allows a simple physical interpretation.  Solutions
to $\bLa$
describe the motion of the perturbation, and $\bLb$ describes the
response of the background system.  Mathur (1990) showed that for any
$\omega$ which is not an eigenvalue of $\bLa$, $\bL$ can inherit a
discrete eigenvalue from $\bLb$.  In this paper, however, we will
consider the eigenvalues of $\bLa$.  Any eigenvalue of $\bLa$
corresponds to an eigenvalue of $\bL$ since the velocity space
structure of $\bLb f_p$ depends solely on the background distribution.
To see this, let $\bLa f_\alpha = \omega f_\alpha$.  We can construct
the response $f_\beta$ such that the total perturbation
$f_p=f_\alpha+f_\beta$.  It satisfies the inhomogeneous equation
\begin{equation}
(\bL-\omega)f_\beta= \sum_{i=1}^3 \pdx{f_b}{v_i}\pdx{\Phi_\alpha}{x_i}
\label{driven}
\end{equation}
Equation \myref{driven} may be singular.  We thus can not simply
invert $\bL-\omega$.  Instead one must establish that at least one
solution $f_\beta$ exists.  This can be seen by going back to the
linearized Vlasov equation \myref{lvlasov}, where we use the full time
dependence to solve a linear hyperbolic problem in a periodically driven
field $e^{i\omega t}\Phi_\alpha$:
\begin{equation}
\dot{f_\gamma}-i\bL f_\gamma = e^{i \omega t}
			\sum_{i=1}^3
\pdx{f_b}{v_i}\pdx{\Phi_\alpha}{x_i}
\label{drivenl}
\end{equation}
Start with the initial value
$f_\gamma(t=0)=0$, and we have a unique solution to the initial value
problem.  Wait for a time $T$, and Fourier transform the result,
yielding $f_\gamma(T,\omega_t)$.  Repeat
this for larger values of $T$, taking the limit as
$T\longrightarrow\infty$.  Since we are working with a stable system,
there can be no growing modes with negative imaginary eigenvalues.
Any damped modes with negative eigenvalues we simply discard.  We take
only the resonant frequency piece, and transform back to get
$f_\delta=f_\gamma(T=\infty,\omega_t=\omega) e^{i \omega t}$.  This will
satisfy~\ref{drivenl}.  To see this, take any interval
$[t_0,t_1=t_0+\omega]$ with $t_0>0$, and apply a discrete Fourier
transform on that interval to equation~\ref{drivenl}.  The RHS
contains only one frequency component $\omega$.  Therefore the LHS
operator $L_d = \partial_t + \bL$ applied to the solution
$f_\epsilon=L_d f_\gamma$ will contain only that same frequency
component.  Since $L_d$ does not depend on $t$, it can only map
functions to zero, but not generate new frequencies.  Thus $f_\delta$
solves~\ref{drivenl}, and $f_\beta=f_\delta e^{-i\omega t}$.

To numerically approximate \myref{driven}, one can discretize the
operators by expanding all functions in a given basis (Pen and Jiang
1992, Pen 1992).  If we expand $f_\beta$ as a discrete sum along the
lines of section~\ref{jeanssec}, we obtain a sequence of
approximations $f_{\beta}^i$ which will converge to $f_{\beta}$.
For each order in the expansion, the differential and integral
operators are finite and discrete, so we can solve \myref{driven}.  We
use the Gram-Schmidt orthogonalization to invert $\bL-\omega$, and
require orthogonality to $f_\alpha$ when we encounter the matrix
singularity.  Specific examples of exact and series solutions will
follow below.

We now only need to consider $\bLa$.  Its solutions describe the
motion of an ensemble of particles in a static field.  A single
periodic orbit corresponds to a localized distribution
($\delta$-function) in phase space, which is a periodic solution, and
thus a discrete sum of eigenstates.  In the case of integrable
systems, one can describe all eigenstates in terms of the action-angle
coordinates.  In general it is possible to have periodic {\it
patterns}, whether or not individual orbits are periodic.  The pattern
need not have the same period as its constituent orbits.  To
illustrate this point, consider a single orbit in an integrable system
with period $T$.  A single particle corresponds to a distribution
function with pattern period $T$, but $n$ particles equispaced
along the angle variable with have period $T/n$.  If we fill the orbit
with a continuum of particles, the pattern will be stationary.  Now
consider a second orbit with a different period.  If the ratio of the
two periods is a rational number, we can construct a pattern with
period equal to any linear sum of the two constituent orbit periods.
Given a set with a continuum parameter of orbital periods, we can, by
judicious choice of orbits and phase angles, construct a pattern with
any period we wish.  In astrophysical situations, such patterns
might be observable.  We will now invert the procedure and try to
construct patterns directly from the Eulerian description of the
distribution function.

To linear order in phase space, the Liouville theorem assures that
these solutions do not diffuse or damp since dynamical friction is a
higher order non-linear phenomenon.  From \myref{ineq} it follows that
a constant mass density perturbation will have more linear behavior if
it is smeared out in space.  One would thus expect that long
wavelength perturbations are dynamically longer lived, which will make
their detection easier observations with limited spatial resolution.

\section{Spherical Potentials}

In general, the eigenmodes of a system must be computed numerically.
For a few systems, however, it is possible to find exact or series
expressions of some modes.  I will show some examples to illustrate
the effects in simple models.  They are radial toy models with
non-singular van-Kampen modes, which implies the existence of initial
conditions which do not Landau damp.  While they do not have a clear
correspondence to three dimensional real galaxies, they are easier to
calculate and may provide some hints about the behaviour of realistic
systems.

Consider simple power law models, where the background distribution
has a power law dependence on the radial coordinate $\rho \propto
r^n$.  The simplest is $n=0$ which is realized in the Einstein sphere,
see for example (Mikahilovskii 1972).  The Einstein sphere is a stable
distribution describing a constant density sphere where all particles
are on tangential (non-radial) orbits, and the distribution function
is isotropic in the tangent plane.  All perturbation orbits which do
not leave the sphere are periodic, and thus all modes have the same
eigenvalue.  This exemplifies the existence of an isolated eigenvalue
in $\bLa$.

For $n=-1$,
the potential gradient $\nabla \Phi_b=F$ is constant.  Now consider purely
radial perturbation orbits.  The linear Vlasov equation
\myref{lvlasov} becomes
\begin{equation}
-i\bLa f_p = v_r \pdx{f_p}{r} - F \pdx{f_p}{v_r} = i\omega f_p.
\end{equation}
Laplace transforming $f_p$ with respect to $r$, such that $f_p=\int
f_k e^{-kr} dk$ allows us to express
the radial eigenmode as
\begin{equation}
f_k = \exp(\frac{-kv^2}{2F} + i \frac{\omega v}{F} - k |r| ).
\end{equation}
This solution is certainly smooth everywhere except at $r=0$.  We
thus need to modify the background potential to allow a smooth
transition.  The $1/r$ density has rapidly divergent mass, so
one must limit the power law
at some radius $R$, which
provides a characteristic scale for $k$.  By using positive and
negative perturbations of the lower harmonics, one can easily construct
non-trivial waves and modes for standing radial density waves.

The zero eigenvalues of isothermal spheres (or any other power law
system) can also be solved.  This scenario is attractive since
many astrophysical objects have such a dependence.
Let $\sigma_b$ be the velocity dispersion of the background material.
Consider an isothermal Maxwellian perturbation with some other
velocity dispersion $\sigma_p$, as might describe an elliptical
galaxy embedded in some halo.  From $\bLa f_a=0$ we obtain the well
known result $\rho_a = r^n,\ \  n=-2 \sigma_b/\sigma_p$.
The background (halo)
response $\rho_b$ to the imbedding is
\begin{equation}
\rho_b=-\frac{2}{n^2+5n+8}\rho_a,
\label{sigma2}
\end{equation}
and the net density fluctuation is $\rho_p=\rho_a+\rho_b$.  In the
case that $\sigma_b=\sigma_p$, one would obtain $n=-2$.  The
background response is equal and opposite to the perturbation and
exactly cancels it.  In this case, no net perturbation actually
happened, except for relabeling $b$ particles to belong to $p$.  But
if we solve the perturbation equation exactly, $\partial_r r^4 \bL
f_p=0$ becomes
\begin{equation}
\pdx{}{r} (r^4 \pdx{\rho_p}{r}) + 2\pdx{r^3\rho_p}{r} + 2r^2\rho_p=0
\end{equation}
which implies
$n=(-5\pm\sqrt{-7})/2$, and we have a nontrivial solution.  This
eigenmode can be obtained from \myref{sigma2} in the limit as $\rho_r
\longrightarrow \infty$.

\section{Jeans Instability}
\label{jeanssec}

The Jeans instability is well understood with known analytic van
Kampen modes.  It is thus a good testbed for general analysis.  The
spatial translation invariance allows an exact series solution in
terms of an infinite sequence of discrete matrix operators on
successively refined basis functions.

Consider a homogeneous isotropic Maxwellian background fluid whose
distribution function is given as $f_b=\rho_b
\exp(-v^2/2\sigma^2)/\sigma\sqrt{2\pi}$.  Then apply the Jeans
swindle and set the potential $\Phi$ of this component to zero.  Let
the Fourier mode $f_p$ be periodic in space with wave number $k$ in
one dimension (say $x$) and constant along the other two ($y,z$) so
the perturbation equation reads
\begin{equation}
\dot{f_p} + i k v f_p - \frac{4i\pi G \rho_b v}{\sqrt{2\pi}\sigma^3 k}
e^{-v^2/s^2} \int dv f_p = 0
\label{pert}
\end{equation}
where $s\equiv\sqrt{2}\sigma$.  Now $\bL=kv - (k_J^2/\sqrt{2\pi}\sigma
k)v\exp(-v^2/s^2)\int dv$ with $k_J^2\equiv 4\pi G\rho_b/\sigma^2$
being the Jeans wavenumber.   Equation \myref{pert} can be solved
as
\begin{eqnarray}
\dot{f_p} &=& i \bL f_p, \nonumber\\
f_p(t) &=& e^{i\bL t} f_p(0).
\label{time}
\end{eqnarray}
All we need to do is find the eigenvalues of $\bL$ and project
the initial condition $f(0)$ onto the eigenstates to obtain the
complete solution.

The simplest case is the free equation, where $G=0$ so we have no
gravity.  Then the eigenmodes of $\bL$ are $f_{v_0}=\delta(v-v_0)$, Dirac
$\delta$-functions which move with frequency $\omega=k v_0$.
The next simplest case are stationary solutions, i.e. distributions
corresponding to zero eigenvalues, $\bL f_p=0$.  Let $f_p =
\exp(ikx) (a \delta(v) + b m(v))$ where
$m(v)=\exp(-v^2/s^2)/s\sqrt{\pi}$ is a Maxwellian
distribution similar to $f_b$.  Equation \myref{pert} becomes
\begin{equation}
\left( \frac{k}{k_J}\right)^2-1=\frac{a}{b}
\label{static}
\end{equation}
and we have solved the static perturbation equation as being the sum
of the Gaussian and a $\delta$ function.  The relevant limits are:
\begin{itemize}
\item{$k\rightarrow\infty$}: for very short wavelength perturbations,
$b\rightarrow 0$ and our solutions are $\delta$-functions, which solve
the free equations as expected.
\item{$k\rightarrow k_J$}: at the Jeans' length, the solution is a
plain Gaussian without any delta function, which is a static
perturbation.
\item{$k\rightarrow \sqrt{2} k_J$}: this is the equality point, where
half the density is in a $\delta$-function and half is in the
Gaussian, so $a=b$.
\item{$k\rightarrow 0$}: it certainly seems curious that we have static
solutions with wavelengths much longer than the Jeans' length.  Note
that $a\rightarrow -b$, which means that $\rho_p\rightarrow 0$ and
the net density fluctuation tends to zero.  All the structure
is in velocity space, implying that we can always have coherent stable
perturbations, even though they contain less and less mass.
\end{itemize}

We can now consider the complete spectral decomposition of $\bL$.  From
the static and free solutions we are led to the ansatz $f_p=
\exp(ikx) (\mu \delta(v-v_0)+ f_h)$ and expand in Hermite polynomials
\begin{equation}
f_h = \sum_{i=0}^\infty c_i N_i H_i(\nu) e^{-\nu^2}
\label{hermite}
\end{equation}

where $\nu\equiv v/s$, and we get $sc_0=\rho_h$.  The
$H_i={1,x,x^2-1,...}$ are defined using the conventions of Gradshteyn
and Ryzhik (1980), and the normalization
$N_i=(2^nn!\sqrt{\pi})^{-1/2}$.  The completeness of this expansion
for $L^1$ Lebesque integrable functions is shown in (Keener).  Note
that projection \myref{hermite} is onto a skew basis, where the
inverse projection occurs through plain Hermite polynomials $c_i=N_i
\int d\nu H_i f_h$ without an exponential weighting.

\newcommand{\bfL}{{\bf L}}
\newcommand{\bfT}{{\bf T}}
\newcommand{\bfV}{{\bf V}}
First, set $\mu=0$.  Substituting \myref{hermite} into \myref{pert},
multiplying by $N_iH_i$ and integrating over $d\nu$,
we have turned the continuum equation \myref{pert} into a discrete
matrix equation, so writing the $c_i$ as a column vector $\vec{c}$, we
obtain $\dot{\vec{c}}=i\bfL\vec{c}$ where $\bfL$ now becomes
\begin{eqnarray}
\bfL&=&2\pi^{1/4}\sigma k\left(\bfV -\frac{k_J^2}{k^2} \bfT\right),
\nonumber\\
\bfT &=& \left(\begin{array}{cccccc}
	0	&0	&0	&0	&0	& \\
	1	&0	&0	&0	&0	& \\
	0	&0	&0	&0	&0	&\cdots \\
	0	&0	&0	&0	&0	& \\
	0	&0	&0	&0	&0	& \\
		&	&\vdots	&	&	&\ddots
	\end{array}\right) \nonumber\\
\bfV &=&  \left(\begin{array}{ccccccc}
	0	&1	&0	&0	&	&	& \\
	1	&0	&\sqrt{2}&0	&\cdots &0	&\cdots\\
	0	&\sqrt{2}&0	&\sqrt{3}& \\
	0	&0	&\sqrt{3}&\ddots&	& \\
		&\vdots	&	&	&0	&\sqrt{n}& \\
		&0	&	&	&\sqrt{n}&0 	& \\
		&\vdots	&	&	&	&	&\ddots
	\end{array}	\right)
\end{eqnarray}
The eigenvalues are easily read off from $\bL$.  If $k>k_J$, we can
symmetrize entries $\bL_{12}$ and $\bL_{21}$ to their geometric mean
(Wilkinson 1965), yielding all real eigenvalues, implying that all
eigenmodes are stable.  The apparent dispersion becomes clear: a
density perturbation $\delta c_0$ projects onto all eigenstates, which
propagate at different velocities.  Only the norm of $\vec{c}$ is
conserved, and $c_0$ clearly decays, explaining the dispersion in
density space.  For $k=k_J$, the matrix becomes explicitly singular,
with the Gaussian $c_0$ being a static eigenmode, as analyzed above.
For $k<k_J$, we obtain two (and only two) imaginary eigenvalues,
$\omega_{1,2}\approx\pm 2i\pi^{1/4}k\sigma \sqrt{k^2/k_J^2-1}$.  This
result holds for small $k$ and is obtained as leading order term from
the continued fraction expansion described below, by directly
observing the singularity of the matrix.  There are only two unstable
eigenstates, one growing and one decaying.

Let us now construct eigenstates.  The eigenstates of $V$ are
$\delta$-functions, as is easily verified from the recurrence relation
of Hermite polynomials.  Unfortunately, convergence to such
discontinuous functions is quite slow.  We thus return to our ansatz
and solve for the analytic part separately.  Restoring $\mu$ into our
calculation, we can solve for the eigenstate iteratively.  Substitute
$\mu\delta(v-v_0)$ back into equation \myref{pert}.  We then obtain
\begin{equation}
\left(\bfL - \frac{kv_0}{\pi^{1/4}} {\bf I}\right) \vec{c} =
\frac{v_0k_J^2}{k}		\left( \begin{array}{c}
			0 			\\
			\mu  	\\
			0 			\\
			0			\\
			\vdots
	       \end{array} \right)
\label{matd}
\end{equation}
which we can solve for $\vec{c}$, unless $k=0$, which we had discussed
earlier.  Gaussian elimination is most easily applied from the bottom
up, and we obtain for the top two coefficients
\begin{eqnarray}
\lambda c_0 +c_1 &=&0 \\
\left(1-\frac{k_J^2}{k^2}\right) c_0+\left(\lambda-C(\lambda^{-1})\right)c_1&=&
	\frac{k_J^2v_0}{2\pi^{1/4}k^2\sigma}\mu
\label{top2}
\end{eqnarray}
Here $C(x)$ is defined from the continued fraction
\begin{equation}
C(x)={2x\over{1-{3x\over 1-{4x\over 1-{5x\over 1-\cdots}}}}}
\end{equation}
while $\lambda=-v_0/2\sigma\pi^{1/4}$.  Knowing the first two
elements, the remainder are simply given by the tridiagonal recurrence
relation \myref{matd}.  This algorithm can be verified by starting with
a finite matrix and iteratively refining the solution.

The same method can be applied to solve for the imaginary eigenvalue,
where we simply require \myref{top2} to be singular, and interpret
$\lambda$ as an eigenvalue.  We thus have an algebraically implicit
expression for the growth factor.  Since the imaginary eigenvalue
is unique up to sign, a growing ansatz should obtain the correct
solution and the implicit solution described here
equivalent to equation (5-31) in Binney and Tremaine (1987).

The qualitative features are easy to extract.  As
$k\rightarrow \infty$ or $v_0 \rightarrow \infty$, the analytic
component vanishes, $\vec{c}\rightarrow 0$ and we recover the free
non-interacting solution.  As $k\rightarrow 0$ we obtain
$-\lambda c_1=c_0=-\mu$ unless we move $v_0$ to $\infty$ at the same
time, in which case arbitrary ratios of $c_0/c_1$ can be achieved.

We can compare this to the analytic expression for the van Kampen mode
\begin{equation}
f_p(v)=\frac{k_J^2}{(2\pi\sigma^2)^{3/2}k^2}
	\frac{k v e^{-v^2/s^2}}{kv-\omega} + \left[ 1-\frac{k_J^2}{k^2}
{\bf \cal W}_V(\frac{\omega}{k\sigma})\right]\delta(v-\omega/k)
\end{equation}
where
\begin{equation}
{\bf \cal W}_V(Z)=\frac{1}{\sqrt{2\pi}}\wp \int_{-\infty}^{\infty}
	\frac{x}{x-Z} e^{-x^2/2} dx
\end{equation}
and $\wp$ denotes the Cauchy principal value for the integral.
We verify correct convergence for $k>k_J$.  The matrix result,
however, is general, and correctly yields all the continuum and
discrete eigenvalues and eigenvectors for all values of $k$.

This example illustrates the separation between the solutions of
$\bLa$ which are simply the $\delta$ functions, and the contribution
from $\bLb$ which contains the response.  This approach can be
numerically extended by expanding the distribution function at every
point in space and solving the more complex set of equations for
arbitrary potentials.

\section{Discussion}

Formally, $\bL$ appears self-adjoint, which unfortunately only holds
for C$^1$ (differentiable) functions.  But as we saw, the
self-adjointness can also be elucidated for quite pathological
function domains.  In non-trivial background potentials, smoothness of
eigenmodes should increase since $\bLa$ contains a $\partial_v$,
which would diverge strongly for singular functions, such as occurred in
the Jeans analysis.

For general systems such as elliptical galaxies, there is no way to
separate the distribution function into the product of a radial and a
velocity piece, as we did in all the examples.  This foils any attempt
to apply a Fourier or other integral transform to $r$ and $v$
separately to express the differential operators as algebraic ones.
Already in the Jeans example, we saw that one cannot define dispersion
measures for van Kampen modes.  In the absence of new tools, the
general problem is very difficult, which is to be expected of partial
differential equations.  Linearity allows systematic numerical sudies
as we will see below.

The assumption of being a small perturbation holds as long as the
inequality \myref{ineq} is satisfied, which is even possible in the
presence of a divergent $\delta$-function.  One need only make sure
that the coefficient of that $\delta$-function is positive to prevent
any possibility of causing negative densities in phase space.

Most elliptical galaxies are observed to possess some weak small scale
structure, which should certainly be subject to perturbative modeling.
If we can have a complete analytic or numerical understanding of the
eigenmodes, these perturbations can teach us much about the detailed
potential structure.  A direct prediction of this analysis is that
perturbations should posses a discrete symmetry: positive and negative
density perturbations should occur with comparable frequencies, and
display similar patterns.  Linearity of \myref{lvlasov} allows us to
reverse the sign of the perturbation $f_p \longrightarrow-f_p$, so we
do not expect small perturbations to be skewed when averaged over many
instantiations, i.e. $<\rho_p>=0$.  Asymmetry only arises as a
non-linear effect for short wavelength or large mass perturbations.

These long-lived oscillations may help explain the presence of shells
in elliptical galaxies.  Quinn (1984) explained these patterns as
transients that arise as an elliptical galaxy accretes a smaller
system, and what we see is the track of a cannibalized dwarf galaxy,
which forms shells through the phase wrapping process.  Hernquist and
Spergel (1992) suggested that these shells can also form through the
merger of two equal mass spiral systems.  The observed concentric
structure of shells implies a large amount of dynamical friction and
fast phase mixing, thus these shells are relatively short lived.
Schweizer and Seitzer (1988) find that a large fraction of all
ellipticals have shells, which implies a high accretion or merger rate
with corresponding cosmological implications.  From \myref{ineq} we
see that a sufficiently long wavelength perturbation does not exhibit
dynamical friction and is subject to perturbative treatment.  The
diffuse remnant may continue to orbit for several dynamical times with
little diffusion or damping.  Furthermore, the periodic motion of
particles in a shell can excite similar modes in the predator galaxy,
which are also long lived.  Therefore, a numerical simulation must
necessarily take these excited modes into account in order to model
such phenomena accurately.

Under certain assumptions, Case and others have proven that the
spectrum of eigenvalues is continuous except for a finite number of
discrete points.  For an Antonov stable system we know that they
cannot contain positive imaginary components.  The matrix analysis
suggests that complex eigenvectors come in complex conjugate pairs, so
we expect all modes to be stable.  Landau damping then only occurs
through the loss of coherence in a superposition of states when
projected onto the density axis.  But any periodic process would
preferentially excite coherent modes, as would be the case in
satellites orbiting about larger galaxies.  These modes can survive
even after the satellite is disrupted or accreted.

Landau's approach using Laplace transforms is an equivalent method to
solve the initial value problem.  Stability here implies the lack of
frequencies with positive imaginary components.  In the Jeans'
argument, the damping occurs despite the existence of coherent van
Kampen modes, since these are singular and thus violate the assumptions
of the Landau analysis.  In this paper we suggested that van Kampen
modes need not in general be singular.  This implies that there
may exist poles in the inverse transform on the real axis, allowing
undamped modes to exist.

The success of the series expansion for the Jeans instability suggests
that this scheme is a practical algorithm to calculate normal
modes in arbitrary potentials.  A systematic search for observable
modes is now possible.  A large density of states near certain
eigenvalues might lead to easily excitable modes.

A numerical eigenmode expansion of the full six dimensional system
should be feasible if we choose the appropriate basis.  Since our
examples all have a basically Gaussian velocity space structure,
together with spherical symmetry and power law radial dependence, the
natural basis should involve Hermite functions in $v$-space, spherical
harmonics for angular dependence and Bessel functions for the radial
piece.  One can envision a p-mode analysis of galaxies similar to
their very successful application to stars.  Observations of both the
surface density and spatially resolved velocity perturbations can in
principle supply us with a three-parameter data set, which we can
compare to numerical calculations.  This should enable us to determine
the low order harmonics.  Given infinite resolution, a three parameter
observation allows us to infer the full three dimensional structure of
the gravitational potential.  From \myref{ineq} one expects that long
wavelength perturbations should be longest lived, since they are the
most linear and less subject to dynamical friction.  This simplifies
life for both the observer and the theorist, since a moderate
resolution will pin down the fundamental modes and low harmonics.

\section{Conclusions}

The picture presented in this article gives a simple physical
interpretation of perturbations about collisionless systems in terms of
eigenmodes.  I have presented some examples of periodic and stable
oscillations in one dimensional collisionless systems, and discussed
general features of these modes, which one can try to find in
elliptical galaxies.

We have explored a little of the structure of perturbations in
velocity space, and saw that it lends itself to simple analysis when
expanded in a suitable way.  The analytic solutions are limited, but
numerical studies can provide a detailed description of fundamental
modes.  An expansion in Hermite polynomials allows an explicit
transformation of the perturbation equations into self-adjoint form
for general $L^1$ integrable functions, which relaxes the standard
differentiability requirement.  This argues that general stationary
systems should also exhibit periodic nondispersive modes.

The linear perturbation solutions of the six dimensional phase space
are vast, and we have seen but a tiny sampling of its rich solution
space.

\section{Acknowledgements}

I wish to thank David N. Spergel and Jeremy Goodman for very helpful
and stimulating discussions, and Joshua Barnes for showing me the
Einstein sphere.  This work was supported by NSF grant ast88-58145.

\end{document}